\documentclass[letterpaper]{article} 
\usepackage{aaai2026}  
\nocopyright
\usepackage{times}  
\usepackage{helvet}  
\usepackage{courier}  
\usepackage[hyphens]{url}  
\usepackage{graphicx} 
\usepackage{natbib}  
\usepackage{caption} 
\usepackage{algorithm}
\usepackage{algorithmic}
\usepackage{amsmath}
\usepackage{booktabs}
\usepackage{makecell} 
\usepackage{array}
\usepackage{booktabs} 
\usepackage{makecell} 
\usepackage{graphicx}
\usepackage{multirow}
\usepackage{caption}
\usepackage[table]{xcolor}
\usepackage{amssymb}
\definecolor{lightblue}{HTML}{E6F0FF} 
\usepackage{newfloat}
\usepackage{listings}
\DeclareCaptionStyle{ruled}{labelfont=normalfont,labelsep=colon,strut=off} 
\floatstyle{ruled}
\newfloat{listing}{tb}{lst}{}
\floatname{listing}{Listing}
\title{DMTrack: Deformable State-Space Modeling for UAV Multi-Object Tracking with Kalman Fusion and Uncertainty-Aware Association}
\author{
    Zenghuang Fu\textsuperscript{\rm 1,2,*}\quad
    Xiaofeng Han\textsuperscript{\rm 1,2,*}\quad
    Mingda Jia\textsuperscript{\rm 1,2}\quad
    Jinming Yang\textsuperscript{\rm 1,2}\quad
    Qi Zeng\textsuperscript{\rm 1,2}\quad
    Muyang Zhang\textsuperscript{\rm 1,2}\quad
    Changwei Wang\textsuperscript{\rm 3,4}\quad
    Weiliang Meng\textsuperscript{\rm 1,2,\dag}
    Xiaopeng Zhang\textsuperscript{\rm 1,2}\quad
}
\affiliations{

  \textsuperscript{\rm 1} The State Key Laboratory of Multimodal Artificial Intelligence Systems, Institute of Automation, Chinese Academy of Sciences \\
  \textsuperscript{\rm 2} School of Artificial Intelligence, University of Chinese Academy of Sciences \\
  \textsuperscript{\rm 3} Key Laboratory of Computing Power Network and Information Security, Ministry of Education; Shandong Computer Science Center, Qilu University of Technology (Shandong Academy of Sciences) \\
  \textsuperscript{\rm 4} Key Laboratory of Computing Power Internet and Service Computing, Shandong Fundamental Research Center for Computer Science \\
}

\usepackage{bibentry}

\begin{document}

\maketitle
\begin{abstract}

Multi-object tracking (MOT) from unmanned aerial vehicles (UAVs) presents unique challenges due to unpredictable object motion, frequent occlusions, and limited appearance cues inherent to aerial viewpoints. These issues are further exacerbated by abrupt UAV movements, leading to unreliable trajectory estimation and identity switches. Conventional motion models, such as Kalman filters or static sequence encoders, often fall short in capturing both linear and non-linear dynamics under such conditions. To tackle these limitations, we propose DMTrack, a deformable motion tracking framework tailored for UAV-based MOT. Our DMTrack introduces three key components: DeformMamba, a deformable state-space predictor that dynamically aggregates historical motion states for adaptive trajectory modeling; MotionGate, a lightweight gating module that fuses Kalman and Mamba predictions based on motion context and uncertainty; and  an uncertainty-aware association strategy that enhances identity preservation by aligning motion trends with prediction confidence. Extensive experiments on the VisDrone-MOT and UAVDT benchmarks demonstrate that our DMTrack achieves state-of-the-art performance in identity consistency and tracking accuracy, particularly under high-speed and non-linear motion. Importantly, our method operates without appearance models and maintains competitive efficiency, highlighting its practicality for robust UAV-based tracking.

\end{abstract}


\section{Introduction}

\begin{figure}[!t]
    \centering
    \includegraphics[width=0.9\linewidth]{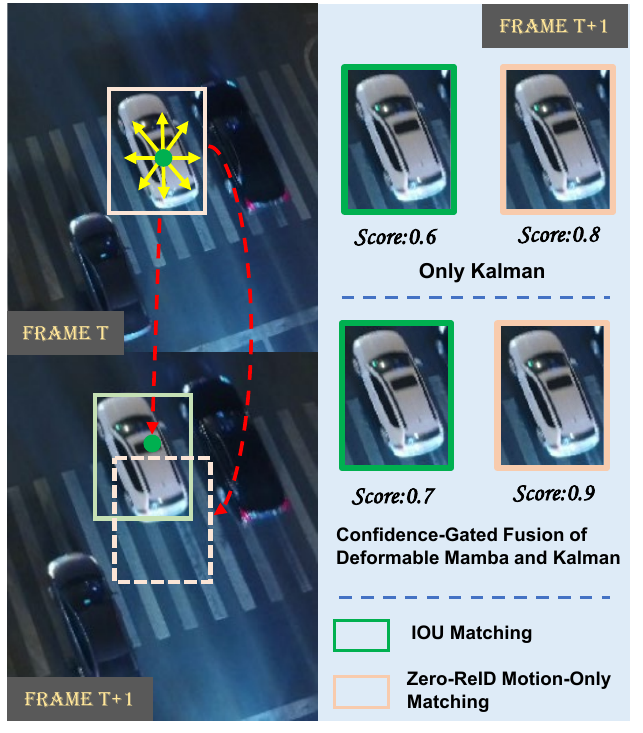}
    \caption{Under abrupt motion (yellow arrows), Kalman-only matching yields lower matching scores (0.6, 0.7), whereas DMTrack achieves higher confidence (0.8, 0.9) through deformable motion modeling, adaptive fusion, and uncertainty-aware association.}
    \label{fig:figure1}
\end{figure}

Multi-object tracking (MOT) is a fundamental task in computer vision with broad applications in surveillance~\cite{yang2023integrating,hassan2024multi}, aerial inspection~\cite{isaac2021unmanned,Dong_2025_CVPR}, intelligent transportation~\cite{gao2024survey,li2024tptrack}, and autonomous systems~\cite{xu2023edge,nagy2025robmot}. In recent years, the increasing use of unmanned aerial vehicles (UAVs) for urban monitoring and emergency response has brought significant attention to UAV-based MOT (UAV-MOT)—a particularly challenging variant of the task~\cite{jiang2021anti}. The top-down view, small target sizes, agile platform movement, and frequent changes in perspective make UAV-MOT uniquely difficult, often resulting in complex, unpredictable object dynamics~\cite{wu2021deep}.

Specifically, tracked objects frequently exhibit high-speed motion, strong non-linearity, repeated occlusion, and disappear-reappear patterns. At the same time, visual cues degrade severely due to low resolution and significant scale variation. These factors collectively undermine the performance of traditional tracking-by-detection frameworks, which depend heavily on high-quality detections and robust appearance-based Re-ID features~\cite{chu2023transmot}. On the one hand, appearance features become unreliable in UAV scenarios due to resolution loss and perspective distortion. On the other hand, standard motion models such as Kalman filters—based~\cite{welch1995introduction} on linear constant-velocity assumptions—struggle with complex trajectories, often leading to inaccurate motion predictions and identity switches.
To mitigate these issues, recent works have explored deep sequence models such as RNNs~\cite{zhang2020long} and neural state-space models to capture long-term motion dependencies~\cite{smith2023convolutional}. Others have proposed learning Kalman-like filters to improve adaptability~\cite{zhou2020tracking,shuai2021siammot}. However, these approaches are typically constrained by linear priors, static update mechanisms, and a lack of dynamic memory control, making them inadequate for modeling highly non-linear and context-dependent motion patterns in real-world UAV settings.

Broadly speaking, existing methods suffer from two critical limitations:(i) They lack the ability to adaptively select motion history based on the current motion context;(ii) They fail to integrate the stability of classical physics-based models (e.g., Kalman filtering) with the modeling capacity of deep neural networks.Figure.~\ref{fig:figure1} clearly illustrates these limitations.Under abrupt motion conditions, traditional Kalman-based matching methods yield lower matching confidence due to rigid linear assumptions. In contrast, our proposed DMTrack successfully resolves this issue, obtaining significantly higher matching scores through deformable modeling and adaptive fusion.

To overcome these challenges, we propose DMTrack—a deformable motion modeling framework specifically designed for UAV-based multi-object tracking. Our DMTrack focuses on improving motion prediction accuracy and robust data association in aerial scenes, while completely removing reliance on appearance features. Our framework comprises three novel components:
\begin{itemize}
\item DeformMamba. A deformable state-space encoder that dynamically predicts temporal offsets to select and interpolate informative motion states from history, enabling flexible modeling of complex, non-linear trajectories.

\item MotionGate. A lightweight gating mechanism that adaptively fuses motion predictions from both Kalman filtering and DeformMamba, leveraging motion context to balance physical stability and learned adaptability.

\item Uncertainty-aware matching strategy. A robust data association method that discards appearance cues and instead leverages motion trend alignment, prediction uncertainty, and temporal consistency to enhance identity preservation during occlusions and reappearances.
\end{itemize}
Extensive experiments on two challenging UAV-MOT benchmarks VisDrone-MOT~\cite{zhu2021detection} and UAVDT~\cite{du2018unmanned} validate that our DMTrack consistently achieves state-of-the-art tracking accuracy and identity consistency, especially under high-speed, non-linear motion and dense crowd conditions. Importantly, our DMTrack maintains real-time performance and does so without any appearance modeling, making it well-suited for real-world UAV applications with constrained computational resources.

\section{Related Works}
\subsection{Tracking-by-Detection Frameworks}



Tracking-by-detection has become the dominant paradigm in multi-object tracking (MOT), where object trajectories are formed by associating detections across frames~\cite{zhang2021fairmot}. Traditional methods often rely on a two-stage pipeline: a detector provides candidate bounding boxes, and an association algorithm links them into trajectories based on motion or appearance cues~\cite{zhang2022bytetrack}. Recent advances such as SORT~\cite{bewley2016simple} and DeepSORT~\cite{wojke2017simple} improve association accuracy by integrating Kalman filters and appearance features. However, in UAV scenarios, appearance cues are unreliable due to low resolution, drastic scale changes, and frequent occlusion~\cite{sandino2020uav}. This limitation has sparked interest in appearance-free trackers, most notably ByteTrack~\cite{zhang2022bytetrack}, which proposes a high-low confidence detection matching strategy solely based on motion predictions from Kalman filtering. Despite its simplicity, ByteTrack achieves strong performance and has inspired several successors including OC-SORT~\cite{cao2023observation} and BoT-SORT~\cite{aharon2022bot}.

Nevertheless, these methods rely heavily on handcrafted motion models (e.g., linear Kalman updates), making them less effective in modeling nonlinear dynamics common in UAV videos, such as sudden acceleration or directional shifts. Our work builds upon this motion-only paradigm, aiming to enhance the trajectory prediction module by incorporating adaptive and learnable motion models.
\begin{figure*}[ht]
    \centering
    \includegraphics[width=0.9\linewidth]{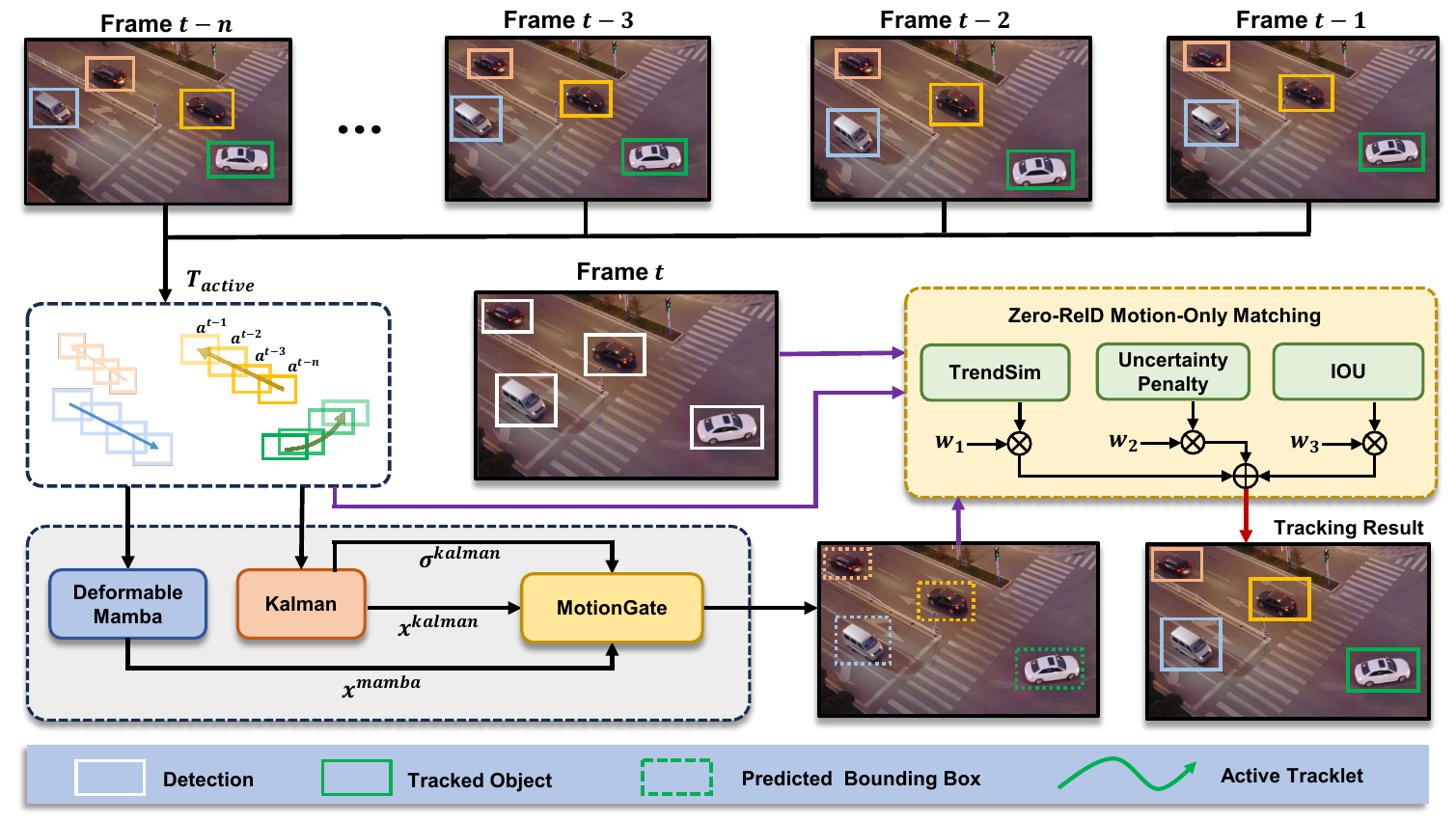}
    \caption{Overview of our DMTrack framework. Our DMTrack first models motion using both Deformable Mamba and Kalman filters, then adaptively fuses their predictions via MotionGate. The final association relies solely on motion cues—IoU, trajectory trend, and uncertainty—to generate robust tracking results without appearance features.}
    \label{fig:Main}
\end{figure*}
\subsection{Motion Modeling in MOT}



Accurate motion modeling is vital for MOT, especially in UAV scenarios where objects exhibit fast and non-linear dynamics~\cite{zheng2025dfa}. Traditional approaches rely on Kalman filters, which assume linear motion and often fail under complex or abrupt trajectories. To enhance adaptability, learning-based models such as RNNs~\cite{zhang2020long} and Transformers~\cite{vaswani2017attention} have been introduced to capture temporal dependencies. While Transformers offer global modeling power, they are often heavy and sensitive to data scale and noise.Recently, Mamba~\cite{gu2023mamba}, a state-space-based sequence model, has shown promise in capturing long-range and non-linear motion patterns efficiently. However, its purely learned structure may lack the inductive bias of classical filters~\cite{karniadakis2021physics}.

To address this, we develop a hybrid scheme combining Kalman filters with Mamba. Our framework leverages DeformMamba for dynamic historical selection and modeling, and introduces MotionGate to fuse Kalman and Mamba predictions based on motion context, ensuring stability and flexibility across scenarios.
\subsection{Re-ID Free Association \& Occlusion Handling}

Re-identification (Re-ID)-free multi-object tracking frameworks have become increasingly popular due to their efficiency and simplicity, especially in scenarios like UAV-based tracking or adverse weather conditions where appearance features are unreliable or unavailable~\cite{bergmann2019tracking,cunico2024multi}. Representative methods such as ByteTrack~\cite{zhang2022bytetrack} and OC-SORT~\cite{cao2023observation} discard appearance embeddings and rely purely on motion cues, using IoU-based bipartite matching with confidence thresholding to achieve robust short-term tracking. While effective under moderate motion and minimal occlusion, these methods often assume linear motion and handle occlusions heuristically, which limits their ability to preserve target identities over long-term occlusion or reappearance events.

To address these challenges, we give an uncertainty-aware matching strategy that discards appearance cues and instead leverages motion trend, prediction uncertainty, and temporal consistency for robust identity association. Unlike prior Re-ID-free approaches that decouple motion and matching, our strategy jointly utilizes deformable modeling and uncertainty-guided fusion to achieve reliable and context-aware tracking in dynamic environments.

\section{Method}
\subsection{Overview}



As shown in Figure~\ref{fig:Main}, we propose Deformable Motion tracking (DMTrack), a lightweight and appearance-free tracking framework tailored for UAV-based multi-object tracking, where non-linear motion and frequent occlusions pose major challenges. Our DMTrack comprises three key modules: DeformMamba, a deformable state-space encoder that adaptively selects and interpolates historical motion states for robust trajectory modeling; MotionGate, a context-aware fusion module that balances Kalman and DeformMamba predictions; and an uncertainty-aware association strategy that ensures reliable identity matching without appearance cues.

Given object detections, our DMTrack predicts future states by combining model-based and data-driven motion cues. Final associations are guided by motion consistency, temporal trends, and prediction uncertainty, enabling robust identity preservation under occlusion and visual degradation. The framework is efficient, flexible, and detector-agnostic, making it well-suited for real-time UAV tracking.

\subsection{Deformable Mamba Module}
\textbf{State Space Model.} State Space Models (SSMs) offer a powerful framework to model temporal dynamics by propagating a hidden state $\mathbf{h}_t$ over time. Formally, an SSM maps input $\mathbf{x}_t$ to output $\mathbf{y}_t$ using the following formulation:

\begin{equation}
\mathbf{h}_t = \hat{A} \mathbf{h}_{t-1} + \hat{B} \mathbf{x}_t,\quad
\mathbf{y}_t = C \mathbf{h}_t + D \mathbf{x}_t.
\end{equation}

To apply SSMs in discrete domains such as language or vision, the continuous-time formulation must be discretized. One common scheme is the zero-order hold (ZOH) method, which approximates a continuous input as piecewise constant over small intervals $\Delta$. Under ZOH, the discretized transition matrices are given by:

\begin{equation}
\begin{aligned}
\hat{A} &= \left(I - \frac{\Delta}{2} A\right)^{-1} \left(I + \frac{\Delta}{2} A\right), \\
\hat{B} &= \left(I - \frac{\Delta}{2} A\right)^{-1} \Delta B.
\end{aligned}
\end{equation}


This discretization allows SSMs to preserve long-range temporal dependencies while remaining computationally tractable. In modern variants such as Mamba, these parameters are further made input-adaptive, allowing dynamic selective scanning and improved temporal modeling across diverse motion patterns.

\begin{figure}[!t]
    \centering
    \includegraphics[width=0.9\linewidth]{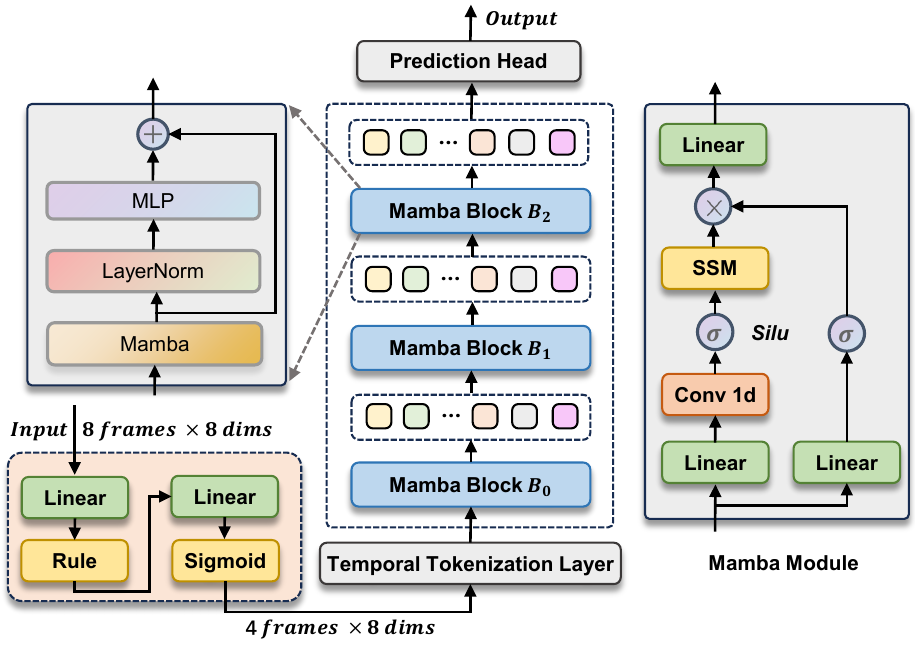}
    \caption{The DeformMamba module.The model predicts temporal offsets to interpolate informative keyframes, tokenizes them, and applies Mamba blocks for trajectory encoding. The final output is generated via a prediction head. Right: internal structure of the Mamba block.}
    \label{fig:mamba}
\end{figure}

\textbf{Deformable Motion Modeling.}To enhance adaptability under complex and non-linear motion patterns, we propose a deformable motion modeling module that selectively attends to motion-relevant frames in the past. 
As illustrated in Figure~\ref{fig:mamba}, the module consists of an offset prediction unit, a keyframe interpolation process, temporal tokenization, a Mamba-based sequence encoder, and a final prediction head.
Each input state is represented as an 8-dimensional vector: $[x, y, a, h, v_x, v_y, v_a, v_h]$, where $(x, y)$ denote the box center, $(a, h)$ denote the box size, and $(v_x, v_y, v_a, v_h)$ represent the corresponding velocities. 
Instead of uniformly aggregating past states, our module predicts continuous temporal offsets and applies differentiable interpolation, allowing the model to dynamically extract useful temporal cues for improved motion forecasting.

\noindent
\textbf{(i) Offset Prediction.}
Given the historical trajectory $\mathbf{X}_{1:T} \in \mathbb{R}^{T \times d}$, where $T=8$ and $d=8$, we flatten it into a 64-dimensional vector and feed it into a two-layer MLP to predict $K$ temporal offsets:
\begin{equation}
\mathbf{O} = \sigma\left( \text{MLP}(\text{Flatten}(\mathbf{X}_{1:T})) \right) \cdot (T - 1),
\end{equation}
where $\sigma(\cdot)$ denotes the sigmoid activation function that constrains each offset $o_i$ within $[0, T-1]$. This design allows the model to flexibly select $K$ informative positions from the past trajectory. The MLP is implemented as:
\begin{equation}
\text{MLP} = \text{Linear}(64, 64) \rightarrow \text{ReLU} \rightarrow \text{Linear}(64, K),
\end{equation}
where $K=4$ is the number of interpolated keyframes. The resulting offset set $\mathbf{O} = \{o_1, o_2, ..., o_K\}$ contains fractional indices that serve as temporal references for keyframe interpolation.


\noindent
\textbf{(ii) Key Frame Interpolation.}
For each predicted offset $o_i$, we determine the two nearest integer time steps $l = \lfloor o_i \rfloor$ and $r = \lceil o_i \rceil$, and interpolate between the corresponding historical states:
\begin{equation}
\tilde{\mathbf{x}}_i = (1 - \alpha) \cdot \mathbf{X}_l + \alpha \cdot \mathbf{X}_r,
\quad \text{where } \alpha = o_i - \lfloor o_i \rfloor.
\end{equation}

\noindent
\textbf{(iii) Temporal Tokenization.}
The interpolated features $\tilde{\mathbf{X}} = \{\tilde{\mathbf{x}}_1, ..., \tilde{\mathbf{x}}_K\} \in \mathbb{R}^{K \times d}$ are linearly projected and normalized to form temporal tokens:
\begin{equation}
\mathbf{Z} = \text{LayerNorm}(\text{Linear}(\tilde{\mathbf{X}})) \in \mathbb{R}^{K \times d'}.
\end{equation}

\noindent
\textbf{(iv) State Representation via Mamba.}
The tokens $\mathbf{Z}$ are processed by a three-layer Mamba encoder to capture long-range temporal dependencies:
\begin{equation}
\hat{\mathbf{z}}_t = \text{MambaEncoder}(\mathbf{Z}).
\end{equation}

\noindent
\textbf{(v) State Prediction Head.}
The final output token $\hat{\mathbf{z}}_t$ is linearly projected to produce the predicted motion state:
\begin{equation}
\hat{\mathbf{x}}_t = \text{Linear}(\hat{\mathbf{z}}_t) \in \mathbb{R}^8.
\end{equation}

This deformable motion modeling process allows the model to dynamically select and interpolate key temporal cues from historical trajectories, enabling it to capture non-linear and abrupt motion patterns often observed in UAV-based tracking scenarios. By explicitly modeling temporal offsets and leveraging Mamba’s efficient sequence encoding, our approach enhances the adaptability and robustness of motion prediction, especially under challenging conditions such as high-speed movement and frequent occlusions.

\subsection{Adaptive Motion Fusion via Confidence-Gated Selection}

\textbf{MotionGate.} In UAV-based multi-object tracking, targets frequently undergo abrupt accelerations and non-linear motion patterns due to the agile flight of aerial platforms. These dynamics make conventional motion predictors prone to identity switches and tracking drift, particularly under occlusion or reappearance. While deep sequence models such as Mamba are capable of learning long-term motion dependencies, their performance may degrade under noise or occlusion, leading to instability in prediction. In contrast, Kalman filters offer robust stability by leveraging physical motion priors, yet they fall short in modeling complex, non-linear trajectories inherent to UAV scenarios.


To bridge this gap, we develop MotionGate, a lightweight and uncertainty-aware fusion module that adaptively combines the outputs of Kalman and Mamba based on motion context and prediction uncertainty. This design is inspired by the complementary strengths of the two models: Kalman filters offer greater robustness in handling short-term, linear motion, while Mamba excels at modeling long-range, complex dynamics. By dynamically balancing their contributions, our MotionGate enhances both the stability and adaptability of motion prediction.

Our MotionGate receives three inputs: the Kalman-predicted state $\mathbf{x}^{\text{kal}}_t \in \mathbb{R}^8$, its associated uncertainty $\boldsymbol{\sigma}^{\text{kal}}_t \in \mathbb{R}^8$, and the Mamba-predicted state $\mathbf{x}^{\text{mam}}_t \in \mathbb{R}^8$. These are concatenated and passed through a shared MLP layer:
\begin{equation}
\begin{aligned}
\mathbf{h} = \text{GeLU}(\text{Linear}_{\text{shared}}([\mathbf{x}^{\text{kal}}_t, \boldsymbol{\sigma}^{\text{kal}}_t, \mathbf{x}^{\text{mam}}_t])) \in \mathbb{R}^{64}.
\end{aligned}
\end{equation}
Then, two branches compute the fusion weight $\boldsymbol{\alpha}_t$ and predicted uncertainty $\boldsymbol{\sigma}^{\text{mam}}_t$:
\begin{equation}
\begin{aligned}
\boldsymbol{\alpha}_t &= \sigma(\text{Linear}_{\alpha}(\mathbf{h})) \in (0, 1)^8, \\
\boldsymbol{\sigma}^{\text{mam}}_t &= \text{Softplus}(\text{Linear}_{\sigma}(\mathbf{h})) + \epsilon.
\end{aligned}
\end{equation}
where $\epsilon$ is a small constant to ensure numerical stability.
The fused state and its uncertainty are then computed as:
\begin{equation}
\begin{aligned}
\mathbf{x}^{\text{fuse}}_t &= \boldsymbol{\alpha}_t \cdot \mathbf{x}^{\text{kal}}_t + (1 - \boldsymbol{\alpha}_t) \cdot \mathbf{x}^{\text{mam}}_t, \\
\boldsymbol{\sigma}^{\text{fuse}}_t &= \boldsymbol{\alpha}_t \cdot \boldsymbol{\sigma}^{\text{kal}}_t + (1 - \boldsymbol{\alpha}_t) \cdot \boldsymbol{\sigma}^{\text{mam}}_t.
\end{aligned}
\end{equation}
Here, each element of the fusion weight $\boldsymbol{\alpha}_t$ represents the per-dimension confidence for the Kalman output, while $(1-\boldsymbol{\alpha}_t)$ weights the Mamba prediction. All vectors are 8-dimensional, representing the bounding box’s center position, size, and velocity.


This fusion mechanism allows fine-grained, per-dimension selection between classical and learned motion models, ensuring smooth and adaptive blending under different motion regimes. 

\begin{figure}[!t]
    \centering
    \includegraphics[width=0.8\linewidth]{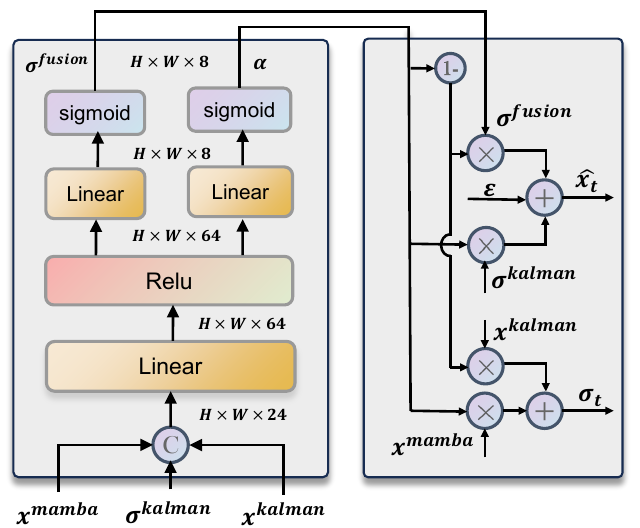}
    \caption{MotionGate module for adaptive motion fusion.computes per-dimension fusion weights from Kalman and Mamba inputs, generating a combined motion state and uncertainty. Left: gating network. Right: weighted fusion process.}
    \label{fig:gating}
\end{figure}
\subsection{Uncertainty-Aware Matching Strategy}
To robustly associate predictions with current-frame detections under motion uncertainty, we give an uncertainty-aware matching strategy that integrates three complementary factors:

\noindent
\textbf{(i) IoU Similarity.}
We compute the standard 2D IoU between the predicted box $\hat{\mathbf{x}}_t = [cx, cy, a, h]$ and the detection box. Both boxes are first converted to $[x_1, y_1, x_2, y_2]$ format before calculating the overlap.

\noindent
\textbf{(ii) Trajectory Trend Similarity.}
To account for motion consistency, we compute the cosine similarity between the predicted trajectory direction and the current motion trend observed from neighboring detections:
\begin{equation}
\text{TrendSim} = \cos(\mathbf{v}_{\text{track}}, \mathbf{v}_{\text{det}}),
\end{equation}
where $\mathbf{v}_{\text{track}} = \hat{\mathbf{x}}_t - \hat{\mathbf{x}}_{t-1}$ and $\mathbf{v}_{\text{det}} = \hat{\mathbf{x}}_t - \mathbf{x}_{\text{det}}$, both using $[cx, cy]$ for direction.

\noindent
\textbf{(iii) Uncertainty Penalty.}
We incorporate a Gaussian penalty to suppress unreliable predictions:
\begin{equation}
\text{Penalty} = \exp\left( - \frac{\|\hat{\mathbf{x}}_{t,\text{pos}} - \mathbf{x}_{\text{det,pos}}\|^2}{\bar{\sigma}^2_{\text{pos}}} \right),
\end{equation}
where $\bar{\sigma}_{\text{pos}} = \frac{1}{2}(\sigma_{cx} + \sigma_{cy})$ is the average positional uncertainty of the predicted state.The final matching score is computed as a weighted sum:
\begin{equation}
    \text{Score} = w_1 \cdot \text{IoU} + w_2 \cdot \text{TrendSim} + w_3 \cdot \text{Penalty},
\end{equation}
where $w_1=0.7$, $w_2=0.2$, and $w_3=0.1$ are empirically chosen to balance spatial alignment, motion consistency, and uncertainty suppression.


This strategy enhances the matching robustness under fast motion or partial occlusions, by emphasizing both geometric and temporal coherence while penalizing unreliable predictions.
To filter out low-quality matches, we set a minimum IoU threshold of 0.3 during data association. Other association settings, including track lifecycle management (e.g., initialization, confirmation, removal), remain consistent with ByteTrack~\cite{zhang2022bytetrack}.

\subsection{Loss Functions for Motion Prediction and Uncertainty Modeling}

To enable precise motion prediction and uncertainty calibration, we jointly train the Deformable Mamba and MotionGate modules using two complementary objectives.

\textbf{State Reconstruction Loss.} We supervise the predicted motion state $\hat{\mathbf{x}}_t$ using ground-truth annotations $\mathbf{x}^{\text{gt}}_t$ with the L1 loss:
\begin{equation}
\mathcal{L}_{\text{state}} = \left\| \hat{\mathbf{x}}_t - \mathbf{x}^{\text{gt}}_t \right\|_1.
\end{equation}
This objective encourages accurate regression of object motion in terms of center, size, and velocity.

\textbf{Confidence-Aware Loss.} To model per-dimension uncertainty, we adopt a negative log-likelihood loss over predicted states:
\begin{equation}
\mathcal{L}_{\text{conf}} = \sum_{i=1}^{8} \left( \frac{1}{2} \cdot \frac{(x_{t,i} - x_{t,i}^{\text{gt}})^2}{\sigma_i^2} + \log \sigma_i \right),
\end{equation}
where $\sigma_i$ is the predicted standard deviation for the $i$-th dimension.

\textbf{Total Loss.} The overall training objective is a weighted sum of the above two terms:
\begin{equation}
\mathcal{L}_{\text{total}} = \lambda_1 \cdot \mathcal{L}_{\text{state}} + \lambda_2 \cdot \mathcal{L}_{\text{conf}},
\end{equation}
where we empirically set $\lambda_1 = 1.0$ and $\lambda_2 = 0.2$ to balance prediction accuracy and uncertainty modeling.

\section{Experiments}
\subsection{Datasets and Metrics}
\textbf{Datasets.} We conduct experiments on two publicly available UAV-based multi-object tracking benchmarks: VisDrone-MOT and UAVDT. Both datasets capture diverse aerial scenes with high object density, frequent occlusions, and motion blur, making them well-suited for evaluating tracking robustness under dynamic motion and degraded visual quality.

VisDrone-MOT comprises four official splits: 56 sequences for training, 7 for validation, 7 for test-dev, and 6 for test-challenge. All sequences are recorded from a drone-mounted camera in urban and suburban environments. The dataset contains ten annotated object categories, including pedestrian, person, car, van, bus, truck, motor, bicycle, tricycle, and awning-tricycle. In our training phase, we utilize all ten classes, while the evaluation phase is limited to five primary categories (car, bus, truck, pedestrian, and van), in accordance with the official evaluation protocol.

UAVDT focuses on vehicle tracking across 50 aerial video clips, including 30 training sequences and 20 test sequences. Objects are annotated with bounding boxes and tracking IDs, and belong to one of three classes: car, truck, and bus. The sequences cover diverse environments such as intersections, highways, and urban streets, captured under varying weather and illumination conditions. During evaluation, we follow the standard protocol and employ VisDrone’s official toolkit to assess performance across all three categories.

\begin{table}[!t]
\centering
\resizebox{\linewidth}{!}{
\begin{tabular}{lccccc}
\toprule
\multirow{2}{*}{Method} & \multicolumn{2}{c}{Visdrone} & \multicolumn{2}{c}{UAVDT} & \multirow{2}{*}{FPS$\uparrow$} \\
\cmidrule(lr){2-3} \cmidrule(lr){4-5}
 & MOTA$\uparrow$ & IDF1$\uparrow$ & MOTA$\uparrow$ & IDF1$\uparrow$ & \\
\midrule
SiamMOT~\cite{shuai2021siammot}    & 31.9 & 48.3 & 39.4 & 61.4 & 11.2 \\
FairMOT~\cite{zhang2021fairmot}     & 34.3 & 46.1 & 41.5 & 59.2 & 17.2 \\
ByteTrack~\cite{zhang2022bytetrack}   & 35.7 & 48.3 & 41.6 & 59.1 & 28.9 \\
UAVMOT~\cite{liu2022multi}        & 36.1 & 51.0 & 46.4 & 67.3 & 12.0 \\
OC-SORT~\cite{cao2023observation}       & 39.6 & 50.4 & 47.5 & 64.9 & 26.4 \\
FOLT~\cite{yao2023folt}          & 42.1 & 56.9 & 48.5 & 68.3 & 29.4 \\
U2MOT~\cite{liu2023uncertainty}         & 42.8 & 53.9 & 47.1 & 65.2 & 24.1 \\
TrackSSM~\cite{hu2024trackssm}      & 41.9 & 55.3 & 48.1 & 65.4 & 24.1 \\
MambaTrack~\cite{xiao2024mambatrack}  & 43.7 & 57.3 & 48.3 & 67.3 & 25.9  \\
MM-Tracker~\cite{yao2025mm}    & 44.7 & 58.3 & 51.4 & 68.9 & \textbf{31.1}  \\
\rowcolor{lightblue}
\textbf{DMTrack(Ours)}          &\textbf{47.8}  &\textbf{61.7}  & \textbf{54.8}  & \textbf{72.2}     &28.4
\\
\bottomrule
\end{tabular}
}
\caption{\textbf{Comparison with state-of-the-art trackers} on the VisDrone and UAVDT test sets.
We report MOTA (\%) and IDF1 (\%) as the main performance indicators, and FPS as the runtime metric.
The best result is highlighted in bold. The $\uparrow$ symbol indicates that higher is better.}
\label{tab:sota_comparison}
\end{table}

\textbf{Metrics.} To comprehensively evaluate tracking performance, we adopt two widely used MOT metrics: \textbf{MOTA}~\cite{bernardin2008evaluating} and \textbf{IDF1}~\cite{ristani2016performance}.
\begin{figure*}[ht]
    \centering
    \includegraphics[width=\linewidth]{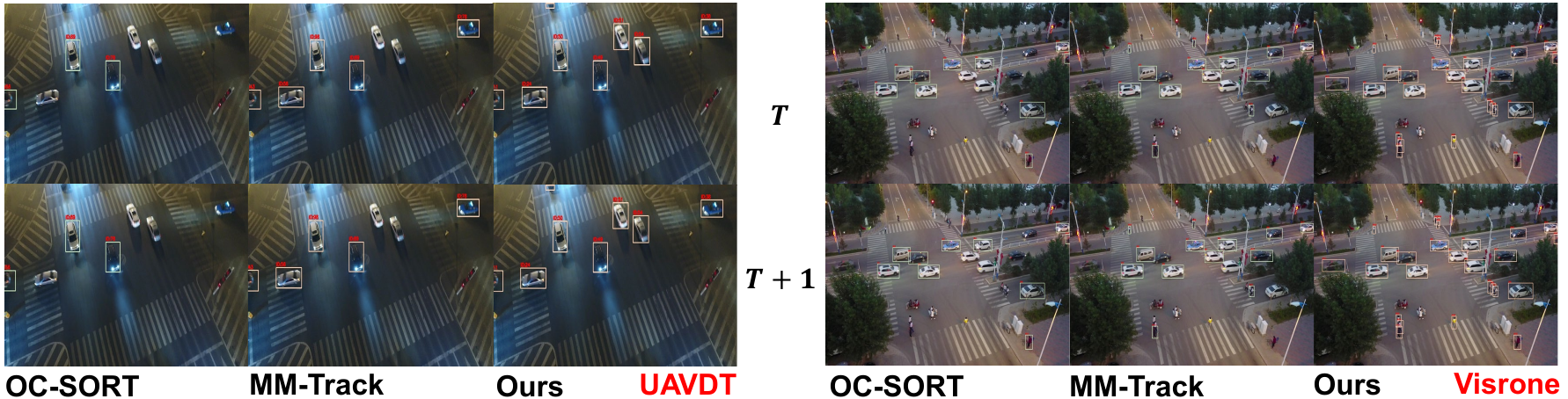}
    \caption{Visual comparison of our DMTrack with OC-SORT and MM-Track.Same numbers indicate consistent identities across frames. Our DMTrack shows better identity preservation than OC-SORT and MM-Track.} 
    \label{fig:experimenta}
\end{figure*}
\subsection{Implementation Details}
\textbf{Detector Training.} We adopt YOLOX-S~\cite{ge2021yolox} as the base object detector for both the VisDrone and UAVDT datasets, following their official train-test splits. The detector is trained with an input resolution of $1280 \times 736$, a batch size of 64, and 60 training epochs. The optimization is performed using the SGD optimizer with an initial learning rate of $1 \times 10^{-4}$. All training was conducted using 4 NVIDIA A6000 GPUs.

\textbf{Tracking Module Training.} We train our tracking module—comprising the Deformable Mamba and MotionGate—separately on each dataset to better capture their domain-specific motion characteristics. One model is trained on the training split of the VisDrone dataset, and another on the UAVDT training set. This per-dataset training strategy accounts for the distinct dynamics and scene layouts of each dataset, such as the faster object movements and broader fields of view in VisDrone, and the more structured traffic patterns observed in UAVDT.

For optimization, we adopt the AdamW optimizer with a learning rate of $1 \times 10^{-4}$ and a batch size of 64. The model is trained for 50 epochs, with a linear warm-up applied during the first 2 epochs.


\begin{table}[!t]
\centering
\resizebox{\linewidth}{!}{
\begin{tabular}{lcccc}
\toprule
\multirow{2}{*}{Motion modeling} & \multicolumn{2}{c}{Visdrone} & \multicolumn{2}{c}{UAVDT} \\
\cmidrule(lr){2-3} \cmidrule(lr){4-5}
 & MOTA$\uparrow$ & IDF1$\uparrow$ & MOTA$\uparrow$ & IDF1$\uparrow$ \\
\midrule
Baseline(KF)             &35.7 &48.3 &41.6 &59.1 \\
LSTM                     &37.5 &49.8 &43.3 &60.2 \\
Mamba                    &39.8 &51.6 &45.5 &62.8 \\
\rowcolor{lightblue}
\textbf{DeformMamba (Ours)} & \textbf{41.8} & \textbf{54.8} & \textbf{46.8} & \textbf{64.5} \\
\bottomrule
\end{tabular}
}
\caption{Compared with Kalman, LSTM, and standard Mamba, our DeformMamba achieves the highest MOTA and IDF1 on both VisDrone and UAVDT, demonstrating superior adaptability to complex motion.}
\label{tab:DM-ablation}
\end{table}

\begin{table}[!t]
\centering
\resizebox{\linewidth}{!}{
\begin{tabular}{lcccc}
\toprule
\multirow{2}{*}{Motion modeling} & \multicolumn{2}{c}{Visdrone} & \multicolumn{2}{c}{UAVDT} \\
\cmidrule(lr){2-3} \cmidrule(lr){4-5}
 & MOTA$\uparrow$ & IDF1$\uparrow$ & MOTA$\uparrow$ & IDF1$\uparrow$ \\
\midrule
Baseline(KF)             &35.7 &48.3 &41.6 &59.1 \\
Mamba-only                   &39.8 &51.6 &45.5 &62.8 \\
Average fusion ($\alpha=0.5$) &42.1 &54.6 &47.5 &65.8 \\
\rowcolor{lightblue}
\textbf{MotionGate (Ours)} & \textbf{44.6} & \textbf{58.2} & \textbf{50.5} & \textbf{68.5} \\
\bottomrule
\end{tabular}
}
\caption{Compared with Mamba-only and average fusion, MotionGate achieves the best MOTA and IDF1 on both datasets, confirming that confidence-gated fusion effectively balances stability and adaptability.}
\label{tab:MG-ablation}
\end{table}

\subsection{Comparison with state-of-the-art}
We compare our proposed DMTrack framework with a range of recent state-of-the-art trackers on the VisDrone and UAVDT test sets. As shown in Table~\ref{tab:sota_comparison}, our method achieves the best overall performance across both benchmarks. Specifically, DMTrack surpasses the previous SSM-based methods—MambaTrack~\cite{xiao2024mambatrack} and TrackSSM~\cite{hu2024trackssm}—by notable margins: on VisDrone, we achieve 47.8\% MOTA and 61.7\% IDF1, outperforming MambaTrack by +4.1\% MOTA and +4.4\% IDF1; on UAVDT, our method obtains 54.8\% MOTA and 72.2\% IDF1, improving over MM-Tracker~\cite{yao2025mm} by +3.4\% MOTA and +3.3\% IDF1. Despite relying solely on motion cues without any appearance features, our method also maintains competitive runtime speed (28.4 FPS), making it highly suitable for real-time deployment.

Figure~\ref{fig:experimenta} further provides visual comparisons across two representative benchmarks. Compared to OC-SORT and MM-Track, our DMTrack generates more accurate and stable trajectories across frames $T$ and $T+1$, especially in cluttered scenes or under abrupt motion. This visual superiority validates the effectiveness of our deformable motion modeling, adaptive fusion, and zero-ReID matching pipeline.

\subsection{Ablation study.}
\textbf{Baseline model.} For comparison, we adopt a strong baseline tracker that integrates YOLOX-S for object detection, an extended Kalman Filter for motion estimation, and a two-stage spatial matching scheme for association.

\textbf{Effect of DeformMamba.} The DeformMamba module is evaluated by comparing it against traditional Kalman filters (KF), LSTM-based predictors, and the standard Mamba encoder. As reported in Table~\ref{tab:DM-ablation} , our method achieves the highest MOTA and IDF1 on both VisDrone and UAVDT, significantly outperforming all baselines. On VisDrone, DeformMamba improves MOTA by +6.1\% and IDF1 by +6.5\% over the Kalman baseline. Similar gains are observed on UAVDT, with +5.2\% MOTA and +5.4\% IDF1. These results highlight the advantage of deformable temporal modeling with adaptive offset selection in handling complex and non-linear motion patterns.

\textbf{Effect of MotionGate Fusion.} We compare MotionGate with two alternatives: using Mamba alone and averaging Kalman-Mamba predictions. As shown in Table~\ref{tab:MG-ablation}, our MotionGate module achieves the best performance across both benchmarks, outperforming average fusion by +2.5\% IDF1 on VisDrone and +2.7\% on UAVDT. This confirms the effectiveness of confidence-gated selection in adapting to diverse motion dynamics.

\textbf{Effect of Uncertainty-Aware Matching.} Uncertainty-Aware Matching improves upon both IoU-only and  $\sigma$-only variants, as shown in Table~\ref{tab:UM-ablation}. By combining motion trend, uncertainty, and spatial alignment, our method achieves better identity preservation on both datasets.
\begin{table}[!t]
\centering
\resizebox{\linewidth}{!}{
\begin{tabular}{lcccc}
\toprule
\multirow{2}{*}{Motion modeling} & \multicolumn{2}{c}{Visdrone} & \multicolumn{2}{c}{UAVDT} \\
\cmidrule(lr){2-3} \cmidrule(lr){4-5}
 & MOTA$\uparrow$ & IDF1$\uparrow$ & MOTA$\uparrow$ & IDF1$\uparrow$ \\
\midrule
Baseline(only-IOU)             &35.7 &48.3 &41.6 &59.1 \\
$\sigma$-only                 &32.3 &46.8 &37.3 &57.2 \\
\rowcolor{lightblue}
\textbf{Uncertainty-Aware Matching(Ours)} & \textbf{40.8} & \textbf{52.8} & \textbf{43.8} & \textbf{63.5} \\
\bottomrule
\end{tabular}
}
\caption{Our method outperforms both IoU-only and $\sigma$-only baselines, showing that combining motion trend, uncertainty, and geometry improves identity preservation.}
\label{tab:UM-ablation}
\end{table}


\textbf{Combined Benefits of All Modules.} Table~\ref{tab:all-ablation} presents the cumulative impact of each proposed module. Beginning with the baseline tracker, we incrementally incorporate DeformMamba, MotionGate, and Uncertainty-Aware Matching. Each module consistently improves both MOTA and IDF1 scores on the VisDrone and UAVDT datasets, highlighting its individual contribution to tracking accuracy and identity preservation. With all components integrated, DMTrack achieves 47.8\% MOTA and 61.7\% IDF1 on VisDrone, and 54.8\% MOTA and 72.2\% IDF1 on UAVDT—marking substantial improvements over the baseline, with average gains of +12.1\% MOTA and +13.4\% IDF1. Importantly, these performance boosts come with minimal computational overhead, as the model maintains real-time efficiency at 28.4 ms per frame, underscoring the practicality of our design.
\begin{table}[!t]
\centering
\resizebox{\linewidth}{!}{
\begin{tabular}{cccc|cc|cc|c}
\toprule
\multicolumn{4}{c|}{\textbf{Modules}} & 
\multicolumn{2}{c|}{\textbf{VisDrone}} & 
\multicolumn{2}{c|}{\textbf{UAVDT}} & 
\multicolumn{1}{c}{\textbf{FPS}} \\
\textbf{B} & \textbf{DM} & \textbf{MG} & \textbf{UAM} & 
\textbf{MOTA↑} & \textbf{IDF1↑} & 
\textbf{MOTA↑} & \textbf{IDF1↑} & 
\textbf{(ms)↑} \\
\midrule
\checkmark & & & & 35.7 & 48.3 & 41.6 & 59.1 & 28.9 \\
\checkmark & \checkmark & & & 41.8 & 54.8 & 46.8 & 64.5 & \textbf{29.8} \\
\checkmark & \checkmark & \checkmark & & 44.6 & 58.2 & 50.5 & 68.5 & 28.5 \\
\rowcolor{lightblue}
\checkmark & \checkmark & \checkmark & \checkmark & \textbf{47.8} & \textbf{61.7} & \textbf{54.8} & \textbf{72.2} & 28.4 \\
\bottomrule
\end{tabular}
}
\caption{Ablation studies on VisDrone and UAVDT test sets. The ↑ means that higher is better. B: Baseline, DM: Deformable Mamba, MG: MotionGate, UAM: Uncertainty-Aware Matching.}
\label{tab:all-ablation}
\end{table}

\section{Conclusion}

We propose DMTrack, a deformable motion tracking framework designed for UAV-based multi-object tracking. To address the challenges of non-linear motion and frequent occlusions, we introduce DeformMamba, which dynamically interpolates motion states using adaptive temporal offsets for improved motion prediction. We further develop the MotionGate, a lightweight fusion module that adaptively combines Kalman and Mamba predictions to balance stability and adaptability. To enable appearance-free data association, we give an uncertainty-aware matching strategy that leverages IoU, motion trends, and prediction uncertainty for reliable identity preservation.

Extensive evaluations on the VisDrone and UAVDT benchmarks demonstrate that our DMTrack achieves state-of-the-art performance in both tracking accuracy and identity consistency, while maintaining real-time efficiency. Our DMTrack offers a practical and robust solution for UAV-MOT and lays the groundwork for future research in adaptive motion modeling and appearance-free association strategies.

\bibliography{aaai2026}

\end{document}